

\magnification=\magstep1
\vsize = 23.5 truecm
\hsize = 15.5 truecm
\hoffset = .2truein
\baselineskip = 14 truept
\nopagenumbers

\def\hef{ $^4$He }

\def\rr { {\bf r } }
\def\rp { {\bf r' } }

\def\bk { {\bf k} }

\topinsert
\vskip 4 truecm
\endinsert

\centerline{\bf DENSITY FUNCTIONAL IN LIQUID HELIUM: RECENT }
\centerline{\bf APPLICATIONS TO DROPLETS AND SURFACES}

\vskip 16 truept

\centerline{\bf Franco Dalfovo, Andrea Lastri, and Sandro Stringari}

\vskip 5 truept

\centerline{Dipartimento di Fisica, Universit\`a di Trento }
\centerline{I-38050 Povo, Trento, Italy }

\vskip 2 truecm


\centerline{\bf 1.  INTRODUCTION}
\vskip 12 truept

In  the present contribution we present  some of the
most  recent results of density functional theory for
inhomogeneous systems of liquid $^4$He. Almost ten years ago, one
of us (S.S.) introduced a simple density functional for liquid
$^4$He and $^3$He. The functional was used to study the
ground state properties of  the  free surface [1] and of
droplets [2]. The theory has been later refined and extended
in order to improve its accuracy and  to make it applicable to
a more microscopic level. Finite range
correlations have been explicitly introduced [3]
to explore strongly  inhomogeneous systems, like layered structures of
liquid helium on solid  substrates [4] and near impurities [5] or
vortices [6]. $^3$He-$^4$He mixtures have been also extensively studied
with the same formalism [7].

It is a good time to see what the situation is. There are two major
questions which have to be addressed. First, what is the present
understanding of the foundations of density functional theory.
For instance, what is the  connection between
phenomenological density functionals and microscopic approaches,
as well as the applicability of the theory to dynamic problems
through a time dependent formulation.
Second, what is the predictive power of the theory and its
accuracy.  In the present contribution we will mainly concentrate
on the second  problem.

We will present  a new density functional [8] (hereafter
called Orsay-Trento functional), which is the result
of a natural evolution and improvement of the theory starting
from the original zero-range  functional of Refs.~[1,2].
We will focus on the key ingredients of the theory and we will
show how they determine important predictions concerning
inhomogeneous systems. These predictions turn out to be
quite accurate even  at the level of interatomic distances.
We will present  a selected set of  results, referring to
Refs.~[8,9] for a more detailed discussion.

\vskip 28 truept

\centerline{\bf 2. DENSITY FUNCTIONAL: EQUILIBRIUM PROPERTIES}
\vskip 12 truept

Let us consider the complex wave function
$$
\Psi (\rr,t) = \Phi(\rr,t) \exp \left({i \over \hbar} S (\rr,t)
\right)  \ \ \ ,
\eqno (1)
$$
where the real function $\Phi$ is related to the one particle density by
$\rho=\Phi^2$, while the phase $S$ is related to the velocity of the
fluid  by ${\bf v} = (1/m) \nabla S$. In the  density functional
approach, the energy  of the   system at zero temperature is assumed
to be a functional of $\Psi$:
$$
     E \  = \   \int \!d\rr \ {\cal H} [\Psi, \Psi^*] \ \ \ .
\eqno (2)
$$
In the calculation of the ground state, only states with zero velocity  are
considered, so that the energy is simply  a functional of the one-body
density $\rho(\rr)$. The ground state is then found by minimizing the
energy with respect to $\rho$.  One usually write the energy as the sum of
the   kinetic energy of non-interacting particles ("quantum pressure")
plus a term   incorporating the effects of quantum  fluctuations and
correlations:
$$
E \ = \ \int \! d\rr \ {\cal H}_0 [\rho] \ = \ \int \!
d\rr  \ {\hbar^2 \over 2m}  ( \nabla \sqrt{\rho})^2  \ +\  E_c[\rho]
\ \ \ .
 \eqno (3)
$$
Minimization with respect to the density leads to the Hartree equation
$$
\left\{ - {\hbar^2 \over 2m} \nabla^2  + U [\rho, \rr] \right\}
\sqrt{\rho(\rr)} \ = \ \mu \sqrt{\rho(\rr)} \ \ \ ,
\eqno (4)
$$
where $U[\rho, \rr] \equiv \delta E_c /\delta \rho(\rr)$ acts as a  mean
field,  while the chemical potential $\mu$ is introduced in order to
ensure the normalization  of the density.

A derivation of functional  ${\cal H}_0$  starting from first principle  is
not available. One  then resorts to approximate schemes for the correlation
energy. The phenomenological density functional  approach consists of
choosing a reasonable {\it ansatz} for  ${\cal H}_0$, in which a few
parameters are fixed to reproduce  known  properties of the liquid.
For example, the functional of Refs.~[1,2] was written in the
form
$$
 E \ = \ \int \! d\rr \ \left[ {\hbar^2 \over 2m} (\nabla \sqrt{\rho})^2
+  {b\over 2} \rho^2 + {c\over 2}
\rho^{2+\gamma} + d (\nabla \rho)^2 \right]  \ \ \ ,
\eqno (5)
$$
where $b,c,\gamma$ and $d$ are phenomenological parameters fixed
to  reproduce the ground state energy, density and compressibility  in the
homogeneous liquid at zero pressure, as well as the free  surface energy.
Non locality effects have been included  in a more adequate   way by
Dupont-Roc et al. [3], who generalized Eq.~(5)
accounting for the finite range of the atom-atom interaction and the
existence of the hard-core.  In that case, the term quadratic in the
density is replaced by a two-body potential energy, containing a
Lennard-Jones interaction, and the term with the power $\gamma$ is taken
to depend on a "coarse grained density", i.e., the density averaged on a
sphere of atomic size. The three parameters of the functional are fixed to
reproduce the bulk equation of state and the energy per particle at
zero pressure.  The most important feature of this approach is that  the
static response function of the liquid turns out to be strongly
$q$-dependent, with a peak at the roton wave length, in qualitative
agreement  with the experimental data [10].
\topinsert
\vskip 10 truecm
\noindent
{\bf Figure 1.} Static response function in liquid $^4$He at zero pressure.
Points: experimental data [10]; dotted line: from
functional of Refs.~[1,2]; dashed line: from Ref.~[3]; solid line: from
the Orsay-Trento functional.
\vskip 28 truept
\endinsert

The static  response  function $\chi(q)$ plays a key role in density
functional theory, since it fixes the response of the system to  static
density  perturbations. It is consequently a quantity which, in principle,
must be exactly accounted for by the theory.  It is easily
calculated from the density functional by taking  the second functional
derivative of the energy in $q$-space. Actually the functional of
Ref.~[3] provides a static response function which underestimates
significantly the height of the peak in the roton region. For this reason
we have corrected that functional, introducing a new non local term
depending on gradients of the density. The new term is fixed to reproduce
the experimental static response function in the  liquid.  The
function ${\cal H}_0$, entering  Eq.~(3), is then written in this way:
$$
\eqalignno{
{\cal H}_0 &= \
{\hbar^2 \over 2m} (\nabla \sqrt{\rho})^2 \ + \
        {1\over 2} \int \!  d\rp \ \rho(\rr) V_l(|\rr-\rp|) \rho(\rp)
	\ + \ {c_2 \over 2} \rho(\rr)  (\bar \rho_{\rr})^2
        \ + \ {c_3  \over 3} \rho(\rr)  (\bar \rho_{\rr})^3
\cr
    	&-    {\hbar^2 \over 4m} \alpha_s \int \! d\rp \ F(| \rr -\rp |)
	       \left(1-{  \rho(\rr) \over \rho_{0s}} \right)
	       \nabla \rho(\rr) \cdot \nabla \rho(\rp)
	       \left(1-{ \rho(\rp) \over \rho_{0s}} \right)
\ \ \ . & (6) \cr
}
$$
The two-body interaction $V_{l}$ is the  Lennard-Jones interatomic
potential, with the standard parameters  $\sigma=2.556$ \AA \ and
$\varepsilon=10.22$ K, screened at short distance  ($V\equiv 0$ for $r<h$,
with $h=2.1903$\AA).  The  weighted density  $\bar \rho$ is the average of
$\rho (\rr)$ over a sphere  of radius $h$.
Thus the two terms containing $\bar \rho$, with the parameters   $c_2
=-2.411857 \times 10^4$ K \AA$^6$ and  $c_3=1.858496 \times 10^6$  K
\AA$^9$,  account phenomenologically for short range  correlations. The
last term corresponds to a non local correction to the kinetic energy.  The
function $F$ is a simple gaussian $F (r)= \pi^{-3/2} \ell^{-3}
\exp(-r^2/\ell^2) $ with $\ell=1$ \AA, while $\alpha_s=54.31$ \AA$^{3}$.
The factors $(1-   \rho /\rho_{0s})$, with $\rho_{0s}=0.04$ \AA$^{-3}$, are
included in order to obtain a pressure dependence of the static response
function close to the one predicted by Diffusion Monte Carlo simulations
[11]. In conclusion, the function ${\cal H}_0$  is such that the theory
reproduces two important features of the liquid,  namely the equation
of state and the static response function. The situation, as concerns
the quantity $\chi(q)$ is shown in Fig.~1, where the experimental data
are compared with the results of three different functionals.   The  $q=0$
limit is fixed by  the  compressibility of the system,  which, in all
cases, is an input of the theory. It ensures the correct  behaviour in
the long wavelength limit and, consequently, the  correct description of
systems characterized  by slow density variations.  The peak  of the
static response  function in the roton region,  $q \simeq 2$ \AA$^{-1}$,
is very  significant  in characterizing  structural properties on the
interatomic  length  scale.

With the functional described above we have studied the properties of the
free surface, droplets and films. Here we restrict ourselves to a few relevant
results for helium droplets, where a detailed comparison between the
predictions of different density functionals with the ones of Monte Carlo
calculations, is possible. In this case the  Hartree equation (4)
has to be solved in spherical geometry. The ground state energy and
density of small droplets  are  shown in Fig.~2 and Fig.~3, respectively,
as a function of the number of atoms.
The density profile of a droplet of 70 atoms
is shown in Fig.~4, together with the predictions of other calculations.
Some relevant comments follow:
\pageinsert
\vskip 9 truecm
\noindent
{\bf Figure 2.}  Energy  per particle
of helium droplets. The results of the present Orsay-Trento functional
(solid line) are compared with the ones of functional of Refs.~[1,2]
(dotted line) and of Ref.~[3] (dashed line),
as well as with Monte Carlo simulations of Ref.~[12] (crosses) and
Ref.~[13] (circles).
\vskip 9.5 truecm
\noindent
{\bf Figure 3.}  Density profile of helium droplets with $N=8$ to $60$.
The density is normalized to the one of the bulk liquid.
\endinsert
\smallskip
\item{$\bullet$}
The predictions of density functional theory  for the energy of
droplets become closer and closer to the results of {\it ab initio}
simulations, as the accuracy of the static response function is
improved.  We note that the three functionals give practically  the same
surface energy, very close to the experimental one,  and the same bulk
equation of state, while they differ in the behaviour of the static
response function. The close agreement between the predictions of the
Orsay-Trento functional and  the ones of {\it ab initio} simulations seems
to indicate that the new non-local kinetic energy term in Eq.~(6)
accounts properly for correlations on the atomic length scale.   The
detailed form of the functional is not crucial; different
parametrizations  give very similar results provided the static
response function in bulk liquid is  reproduced.
\item{$\bullet$} The density profiles exhibit small regular oscillations.
These oscillations were not found with previous functionals, because the
latter underestimated the height of $\chi(q)$  at the roton wave length. The
connection between the oscillations of the density profile and the
behaviour of the static response function was suggested twenty years ago by
Regge and Rasetti [14]. Subsequent  microscopic calculations did
not provide any conclusive evidence of this effect. The results of
the density functional approach support the Regge's idea. These
oscillations have a regular behaviour when one goes  from small to
very large clusters, up to limit of a
planar free surface. The density oscillations are found to affect also the
the evaporation energy, $[E(N-1)-E(N)]$,  characterizing the mass
distribution of droplets in experimental beams: deviations of the order of
$0.1$ K with respect to a pure liquid drop model are predicted.  These
deviations are signatures of a shell structure ("soft sphere close
packing") induced by quantum correlation at the liquid surface.
\item{$\bullet$} Density oscillations have been recently found in
Diffusion Monte Carlo (DMC) calculations on small droplets by Chin and
Krotscheck [15], even if their results are probably affected by  large
statistical errors. The same authors have more recently  employed a
variational (HNC) method, and found oscillations closer to the ours [16].
A very good agreement with the density functional results has been very
recently  obtained  by Barnett and Whaley in their most recent DMC
calculations [17].
\smallskip
\topinsert
\vskip 11 truecm
\noindent
{\bf Figure 4.}   Density profile of a droplet with $70$ \hef atoms.
Solid line: present work; triangle: DMC simulations of Ref.~[15];
 dashed line: Variational (HNC) calculations [16]; circles: most recent
DMC calculations by Barnett and Whaley [17].
\vskip 28 truept
\endinsert

In other systems, like the free surface, films, porous media, doped
clusters, less information are available from {\it ab initio} methods.
The accuracy of the density functional approach is expected to be similar
to the one shown for helium droplets. We emphasize again that the
freedom in the choice of the parametrization of the starting functional
is only a minor source of indetermination in the predictive power of the
theory, provided the crucial ingredients (equation of state and static
response function in bulk liquid) are properly accounted for.

\vfill\eject

\centerline{\bf 3. DENSITY FUNCTIONAL: EXCITED STATES}
\vskip 12 truept

Starting again from functional (2) one can develop a time dependent
density functional theory in order to study the dynamics of inhomogeneous
systems. In this case one has to apply the least action principle
$$
\delta \int_{t_1}^{t_2} dt \int \! d\rr \left[ {\cal H} [\Psi^*, \Psi]
-\mu \Psi^* \Psi - \Psi^* i \hbar {\partial \Psi \over \partial t }
\right] \ =\ 0
\ \ \ .
\eqno (7)
$$
 The equations of motion for the excited states of the fluid can be  derived
by making variations with respect to $\Psi$ or $\Psi^*$. One finds  a
Schr\"odinger-like equation of the form
$$
(\tilde{H} - \mu )  \Psi = i \hbar {\partial \over \partial t } \Psi \ \ \ ,
\eqno (8)
$$
where $\tilde{H} = \delta E/ \delta \Psi^*$ is an effective
Hamiltonian.  If one looks for linearized solutions
$$
\Psi ({\bf r},t) = \Psi_0({\bf r}) + \delta \Psi({\bf r},t)
\eqno (9)
$$
the Hamiltonian  $\tilde{ H}$ takes the form
$ \tilde{H} = \tilde{H}_0 + \delta \tilde{H}$, where the static
Hamiltonian $ \tilde{H}_0  $ fixes the equilibrium state as in
Eq.~(4).   The term  $\delta \tilde{H}$ is linear in $\delta \Psi$ and
accounts  for changes in the Hamiltonian induced by the collective
motion of the system. The Schr\"odinger equation (8) has to be
solved using a self-consistent procedure. This linearized theory concides
with the Random Phase Approximation for bosons. More detailed aspects of
the theory were discussed by  S. Stringari in the same Workshop.

To solve the equations of motion of the time dependent density functional
theory one has  to consider explicitly the velocity dependence of the
function ${\cal H}$. We write
$$
{\cal H} = {\cal H}_0 [\rho] + {\cal H}_v [\rho,{\bf v}] \ \ \  ,
\eqno (10)
$$
where ${\cal H}_0$ is the velocity independent function given in Eq.~(6),
while ${\cal H}_v$ is taken to be [8]
$$
{\cal H}_v = {m \over 2} \rho(\rr) |{\bf v}({\bf r})|^2
             - {m \over 4} \int \! d\rp \ V_J (|\rr-\rp|) \ \rho(\rr)
	     \rho(\rp) \  \left[ {\bf v}(\rr) -{\bf v}(\rp) \right]^2 \ \ \ .
\eqno (11)
$$
The first term is the usual kinetic energy for free particles. The second
term plays the role of a non local kinetic energy, accounting for
backflow effects through an effective current-current interaction
$V_J$.    A similar functional was introduced long time ago by Thouless
[18] to study the flow of a dense superfluid. We fix $V_J(r)$ using  the
phonon-roton dispersion in bulk liquid as input.  We have chosen the
simple  parametrization
$$ V_J (r) = (\gamma_{11} + \gamma_{12} r^2)
\exp(-\alpha_1 r^2)
         + (\gamma_{21} + \gamma_{22} r^2) \exp(-\alpha_2 r^2) \ \ \ ,
\eqno (12)
$$
where the parameters are given in Table I.  With this choice the
phonon-roton spectrum of the uniform liquid at zero pressure is well
reproduced, and its pressure dependence also agrees with the
experimental data. The theory is then used to predict dynamic properties
of non uniform systems.

\topinsert
\centerline{\bf{Table 1}}
\vskip 12 truept
\noindent
Values of the parameters used in $V_J(r)$.
\vskip 2 truecm
\endinsert

We show here what happens in the case of the free surface. The time
dependent part of the wave function (9) has to be expanded in plane waves
in the direction parallel to the surface. The equations of motion are
solved for the $z$-dependent coefficients, where $z$ is the orthogonal
coordinate. In practice, we expand the solutions on a basis of
eigenfunctions of the static Hamiltonian, in order to write the equations
in the form of a matrix diagonalization. One gets the transition density,
the energy and the strength associated with each excited state, for any
given value of the parallel wave vector $k$. The same procedure can be
easily applied to the case of helium films, by adding the external
potential of the substrate.

In Fig.~5 we show the  excitation spectrum of the free surface. Up to
$k \simeq 1.15$ \AA$^{-1}$,  the state with lowest energy is localized at
the surface. In the long wave length limit $k \to 0$ its dispersion
coincides with the hydrodynamic dispersion of ripplons (dotted line).
It deviates significantly from the hydrodynamic law when approaching the
threshold energy $\Delta=8.7$ K for rotons.   A similar deviation has been
found in the dispersion of surface modes in helium films measured in
neutron scattering experiments [19] (points with error bars).  The bulk
phonon-roton dispersion (upper solid line) is also plotted, together with
the experimental data.  We stress again  that, while  the phonon-roton
dispersion in the bulk is used as input to fix the   velocity  dependent
part of the functional, the theory becomes predictive for the dynamic
properties of inhomogeneous systems, like the free surface.
\smallskip
\topinsert
\vskip 10.5 truecm
\noindent
{\bf Figure 5.}  Excitation spectrum of the free surface (see text).
\vskip 28 truept
\endinsert

Below $\Delta$ the surface mode is undamped, while above $\Delta$ it
couples with the continuum of bulk modes (rotons with negative and
positive group velocity) propagating at different angles ($q_z \neq 0$).
This results in a spreading of the strength associated with the
surface mode. Actually the spreading predicted by our theory is rather
small. This can be seen by plotting the dynamic structure function $S(k,
\omega)$, defined by
$$
S(k,q_z,\omega) =  \sum_b \left|  \left(\hat{\rho}^{\dag}_{\bf q}
\right)_{\bk,b} \right|^2  \delta(\omega - \omega_{\bk,b}) \ \ \ .
\eqno (13)
$$
In Fig.~6 (top) we show the dynamic structure function for scattering at
grazing angle ($q_z=0$) on a slab (liquid  between two parallel surfaces at
a distance of $50$ \AA).  For graphical convenience we have replaced the
$\delta$-functions with gaussians of width $0.4$ K. The strength  of the
surface mode is well localized also above $\Delta$, even though  the
strength is partially distributed among bulk modes coupled to  ripplons.
The position of the peak above $\Delta$ is shown also in Fig.~5 as a dashed
line. The hybridization mechanism between ripplons and rotons is
discussed on a general ground in Ref.~[9].
\smallskip
\pageinsert
\vskip 19 truecm
\noindent
{\bf Figure 6.}  Dynamic structure function for a  $50$ \AA\ thick slab
(top) and a film on Sodium (bottom), with coverage of $0.23$ \AA$^{-2}$.
\endinsert

The form of the spectrum can change significantly in liquid helium
films on solid substrates, where the substrate-helium potential produces a
layering of the helium density and the bulk modes are discretized by the
finite size of the system along $z$. An example is given in Fig.~6
(bottom), where we show the dynamic structure function for a film on
Sodium. The areal density of helium is $0.23$ \AA$^{-2}$, which
corresponds to approximately $4$ liquid layers. The surface mode is clearly
visible. Near the roton minimum one notes a new structure coming from the
first liquid layer close to the substrate. It is interpreted as a 2D roton,
whose energy becomes significantly lower than $\Delta$ in the case of more
attractive substrates, reflecting the tendency to solidification.  Similar
structures have been also seen in the experimental spectrum [19].

Finally, we note that the density functional method is also suitable to
investigating the properties of reflection and evaporation of bulk
excitations at the surface. Liquid helium is very peculiar from this
viewpoint. For instance, one can study the one-to-one process of a
roton impinging the surface and ejecting  an atom (quantum evaporation).
Experimental data about evaporation and condensation rates are now
becoming available [20]. The threshold for quantum evaporation at zero
temperature is shown in Fig.~5 (dot-dashed line). The $q=0$ value is the
chemical potential $|\mu|=7.15$ K. Above this curve the solutions of the
time dependent density functional theory are combinations of free atom
states, outside the liquid, and bulk states (phonons and rotons).
{}From their coupling one can
extract the evaporation and condensation probabilities. This work is in
progress.

\vskip 28 truept
\centerline{\bf 4. CONCLUSIONS}
\vskip 12 truept

In this contribution we have briefly traced the evolution of
phenomenological density functional theories, from the one
of Refs.~[1,2] to the most recent Orsay-Trento functional [8]. We
have emphasized the crucial role of the static response function of the
bulk liquid;  together with the equation of state, this quantity has
to be considered
the main ingredient of the density functional apprach. We have shown how
the predictions of the theory improve when the static response function
is properly reproduced. For the time dependent version of the theory one can
introduce backflow effects through an effective current-current interaction.
In that case the phonon-roton dispersion in the uniform liquid is used as
input.  This makes the theory quantitative in the prediction of the excited
states of inhomogeneous systems. We have shown results for the dispersion
of ripplons on the free surface and for the excitations of films. An accurate
description of the excitations in bulk is particularly important to
study the ripplon-roton hybridization mechanism [9] and the process of
quantum evaporation.

The predictions of the theory compare well with available results
of Monte Carlo similations in small systems. The density functional
approach can be  easily applied to large systems too, and consequently
can  be considered a reliable tool to  exploring  the  properties of
inhomogeneous liquid  helium in a rather systematic way.

\vskip 28 truept
\centerline{\bf ACKNOWLEDGMENTS}
\vskip 12 truept

Most part of the results presented in this work comes from a fruitful
collaboration with L. Pitaevskii, L. Pricaupenko and J. Treiner.
This work was
partially supported by European Community grant ERBCHRXCT920075.

\vskip 28 truept

\centerline{\bf REFERENCES}
\vskip 12 truept

\item{[1]}  S. Stringari and J. Treiner,  Phys. Rev. B {\bf
36}, 8369 (1987)

\item{[2]} S. Stringari and J. Treiner, J. Chem. Phys. {\bf 87},
5021 (1987)

\item{[3]} J. Dupont-Roc, M. Himbert, N. Pavloff, and J. Treiner,
J. Low Temp. Phys. {\bf 81}, 31 (1990)

\item{[4]}   E. Cheng, M. W. Cole, W. F. Saam, and J. Treiner,
Phys.  Rev. Lett. {\bf 67}, 1007 (1991); Phys. Rev. B {\bf 46}, 13967
(1992); E.  Cheng, W.F. Saam, M.W. Cole, and J. Treiner, J. Low Temp. Phys.
{\bf 92}, 11 (1993);  L. Pricaupenko and J. Treiner, J. Low Temp.  Phys.
{\bf 96}, 19 (1994)

\item{[5]} N. Pavloff, Thesis, Orsay, unpublished;  F. Dalfovo, Z. Phys. D
{\bf 29}, 61 (1994);  M. Barranco and E.S. Hernandez, Phys. Rev. B {\bf 49}
12078 (1994); F. Ancillotto, E. Cheng, M. W. Cole, and F. Toigo, Z. Phys. B
(in press)

\item{[6]} F. Dalfovo, Phys. Rev. B {\bf 46}, 5482 (1992); F.
Dalfovo, G. Renversez, and J. Treiner, J. Low Temp. Phys. {\bf 89},
425 (1992)

\item{[7]} F. Dalfovo, Z. Phys. D {\bf 14}, 263 (1989); N. Pavloff and J.
Treiner, J. Low. Temp. Phys. {\bf 83}, 331 (1991); J. Treiner, J. Low
Temp. Phys. {\bf 92}, 1 (1993); L. Pricaupenko and J. Treiner, preprint,
Orsay (1994)

\item{[8]} F. Dalfovo, A. Lastri, L. Pricaupenko, S. Stringari, and J.
Treiner, in preparation

\item{[9]} A. Lastri, F. Dalfovo, L. Pitaevskii, and S.Stringari, J. Low
Temp. Phys., in press

\item{[10]}  R.A. Cowley and A.D.B. Woods, Can. J. Phys. {\bf 49},
177 (1971); A.D.B Woods and R.A. Cowley, Rep. Prog. Phys. {\bf 36}, 1135
(1973)

\item{[11]}  S. Moroni, D. M. Ceperley, and G. Senatore, Phys. Rev.
Lett. {\bf 69}, 1837 (1992); S. Moroni, private communication

\item{[12]} R. Melzer and J. G. Zabolitzky, J. Phys. A: Math. Gen.
{\bf 17}, L565 (1984)

\item{[13]} S.A. Chin and E. Krotscheck, Phys. Rev. B
{\bf 45}, 852 (1992)

\item{[14]} T. Regge, J. Low. Temp. Phys. {\bf 9}, 123 (1972); M. Rasetti
and T. Regge, in {\it Quantum Liquids}, edited by J. Ruvalds and T. Regge
(North Holland, 1978) p.227

\item{[15]}  S.A. Chin and E. Krotscheck, Phys. Rev. B
{\bf 45}, 852 (1992)

\item{[16]}  S.A. Chin and E. Krotscheck, preprint (1994)

\item{[17]} R.N. Barnett and K.B. Whaley, private communication

\item{[18]}  D. J. Thouless, Ann. of Phys.  {\bf 52}, 403, (1969)

\item{[19]}  H.J. Lauter, H. Godfrin, V.L.P. Frank and
P. Leiderer in {\it Excitations in Two-Dimensional and Three-
Dimensional Quantum Fluids}, eds. A.F.G. Wyatt and H.J. Lauter (Exeter
1990),  NATO ASI Series B Vol. 257, p.419; H.J. Lauter, H. Godfrin, and
P. Leiderer, J. Low Temp. Phys. {\bf 87}, 425 (1992)

\item{[20]}  A.F.G. Wyatt, J. Low Temp. Phys. {\bf 87}, 453
(1992), and references therein

\end